\documentclass[aps,pra,twocolumn,showpacs,showkeys,preprintnumbers,amsmath]{revtex4-1}

\usepackage{bbm}
\usepackage{esint}
\usepackage{graphicx}
\usepackage[colorlinks]{hyperref}
\hypersetup{linkcolor=[rgb]{0.2,0.4,0.7}} 

\newcommand{\pd}{\partial}
\newcommand{\bs}{\boldsymbol}
\def\res{\mathop{\text{Res}}\limits}

\renewcommand{\d}{\mathrm{d}} 
\newcommand{\ii}{\mathrm{i}}
\let\oldpi\pi
\renewcommand{\pi}{\mathrm{\oldpi}}
\newcommand{\e}{\mathrm{e}}

\begin{document}

\title{Rutherford scattering of electron vortices}

\author{Ruben Van Boxem}
 \email{ruben.vanboxem@uantwerpen.be}
\author{Bart Partoens}
\author{Johan Verbeeck}
\affiliation{EMAT \& CMT, University of Antwerp,\\Groenenborgerlaan 171, 2020 Antwerpen}

\date{\today}

\begin{abstract}
By considering a cylindrically symmetric generalization of a plane wave, the first Born approximation of screened Coulomb scattering unfolds two new dimensions in the scattering problem: transverse momentum and orbital angular momentum of the incoming beam.
In this paper, the elastic Coulomb scattering amplitude is calculated analytically for incoming Bessel beams.
This reveals novel features occurring for wide angle scattering, and quantitative insights for small angle vortex scattering.
The result successfully generalizes the well known Rutherford formula, incorporating transverse and orbital angular momentum into the formalism. 
\end{abstract}

\maketitle

\section{Introduction}

Since the first theoretical description of optical vortices by Nye and Berry~\cite{Nye} and a seminal work by Allen \textit{et al.}~\cite{Allen} describing the role of the orbital angular momentum (OAM), a range of applications has sprouted from topologically distinct optical beams, including particle micro-manipulation~\cite{He,Padgett}, atomic excitation~\cite{Picon}, imaging~\cite{Furhapter,Hemo}, and even quantum information~\cite{Fickler}.
The field has undergone a rapid evolution where focus has shifted quickly from theoretical study to more practical applications.
This rapid evolution is in part due to the widespread availability of coherent light sources and precise optical elements of all shapes and sizes.
Phase vortices, characterized by a local phase singularity of the form $\e^{\ii \ell \varphi}$ in which $\ell$ is an integer, appear not only in optics, but also in acoustics~\cite{Demore} and quantum mechanics~\cite{Zambrini}, where matter waves are seen to form singularities in the same way as light.

Only recently described~\cite{Bliokh_semiclassical} and demonstrated~\cite{Uchida,Verbeeck}, electron vortex beams are a new field of research with promising applications for the characterization of magnetic materials and nano-manipulation~\cite{Verbeeck_tweezers}.
Using many of the concepts as in optics, electron vortex beams can be easily produced.
These procedures include electromagnetic phase plates~\cite{Uchida, Beche}, holographic apertures~\cite{McMorran,Schattschneider_fork,Verbeeck,Verbeeck_spiral}, mode conversion~\cite{Schattschneider_mode_conversion}, and more advanced methods, which rely on the aberration correctors in an electron microscope to manipulate the phase of the wave function directly~\cite{Clark}.
The complex interplay between spin and OAM make electron vortices an interesting topic~\cite{Bliokh_spin_orbit,VanBoxem,Leary}.
The interaction with matter is what makes electron vortex beams useful as a tool to study material properties at a nanoscale~\cite{Verbeeck,Verbeeck_tweezers}.
For example, electron vortex beams can interact inelastically to excite atomic states forbidden by the plane wave selection rules, or elastically resulting in for example OAM-dependent propagation dynamics in the presence of a magnetic field~\cite{Bliokh_magnetic}.
To give a proper description of scattering of vortex waves in this paper we extend historical results to incorporate transverse momentum and OAM into scattering theory.
Our current focus lies on elastic scattering.

In the following, the exact analytical scattering amplitude in the first Born approximation of an electron beam with transverse momentum is presented.
This theory includes all focused beam profiles, including electron vortex beams, as the Bessel beam basis we employ is a complete basis of the Schr\"odinger equation in free space.
The Born approximation is commonly used as a first approximation to both elastic and inelastic scattering phenomena, providing a good result for a wide range of experiments and interactions.
It also forms the basis for treatments of inelastic scattering and more complex quantitative formulas.
The obtained expression is then analyzed in detail for beams with and without OAM.
Interesting features are discussed in both cases.
The influence of misalignment, which is not present in the plane wave case, is studied using properties of the Bessel functions and associated scattering amplitude.
Using these analytical results, a proposal for a new imaging method is explained, which relies on the vortex' dark spot.

This work is structured as follows.
In the next section the quantum theory of scattering is explained, to allow a clear point of insertion for an incoming wave with OAM and transverse momentum.
The third section introduces vortex waves as general solutions of the Schr\"odinger equation in cylindrical coordinates, and presents a vortex counterpart of a plane wave, into which localized physical beam states can be decomposed.
The fourth section contains the calculation of the first-order scattering amplitude for an incoming vortex wave on a Coulomb potential, where the beam is perfectly aligned to the scattering center.
The fifth section contains an experimental proposal to demonstrate the potential application of this theory.
Lastly, the concept of off-center scattering is introduced using the Bessel beam basis and the effects of the off-centering on the previous result is discussed.
\\

\subsection*{Notation}

In order to remove clutter from crucial steps in the calculation, strict notational conventions will be defined here.
Real space coordinates are denoted $\bs r = (x,y,z) = (r,\varphi,z)$.
Basis vectors are written as $\bs e_i$ where $i$ is the relevant coordinate.
Momentum space coordinates are denoted $\bs k = (k_x, k_y, k_z) = (k_\perp, \phi, k_z)$.
It is often advantageous to split any 3D vector $\bs v$ into a $z$-component $v_z$ and a perpendicular component $\bs v_\perp = (v_x, v_y)$ with size $v_\perp$.
Primed variables are denoted $\bs r^\prime = (x^\prime, y^\prime, z^\prime) = (r^\prime, \varphi^\prime, z^\prime)$ and similar for $\bs k^\prime$.
Momentum transfer is denoted $\bs q$ and is equal to $\bs k - \bs k^\prime$.

The usual constants appear in the equations, including the reduced Planck constant $\hbar=h/(2\pi)$ and the electron mass $m_e$.

\section{Quantum scattering on a Coulomb potential: the Rutherford formula}

In this section, the basic quantum scattering formalism will be reviewed, and applied to calculate the Rutherford scattering amplitude.
We begin with the Lippmann-Schwinger equation (considering only the outgoing scattered wave of the two formal scattering wave functions $\Psi^\pm$)~\cite{Ballentine},
\begin{equation} \label{eq:lippmann_schwinger}
 |\Psi \rangle = |\Phi\rangle + \frac{1}{E - H_0 + \ii\varepsilon} V |\Psi \rangle.
\end{equation}
This describes the perturbation caused by a potential, $V$, on the unperturbed (incoming) states $|\Phi\rangle$ which satisfy $H_0 |\Phi\rangle = E |\Phi\rangle$.
As is well-known, one obtains a perturbation series in $V$ by iteratively replacing $|\Psi\rangle$ on the right-hand side of eq.~\eqref{eq:lippmann_schwinger} by the right-hand side of that equation:
\begin{align} \label{eq:born_series}
 |\Psi\rangle\ &= |\Phi\rangle + \frac{1}{E-H_0+\ii \varepsilon} V|\Phi\rangle \notag \\
 &+ \frac{1}{E-H_0+\ii \varepsilon} V\frac{1}{E-H_0+\ii \varepsilon} V|\Phi\rangle + \hdots
\end{align}
In integral form, a simple and intuitive expression can be found for the asymptotical wave function (for large distances $\bs r$ away from the scattering center, and if $V(\bs r)$ is a short-range potential):
\begin{equation} \label{eq:scattering_wave_function}
\Psi(\bs r) = \Phi(\bs r) + f[\Phi,\Psi] \frac{\e^{\ii k r}}{r},
\end{equation}
where the scattering amplitude $f$ is defined from the series in eq.~\eqref{eq:born_series}:
\begin{align}
 f[\Phi,\Psi] &= - \frac{1}{4\pi} \frac{2m_e}{\hbar^2} \int \d^3 \bs r~\e^{-\ii \bs k^\prime \cdot \bs r} V(\bs r) \Psi(\bs r) \notag \\
  &= - \frac{N}{4\pi} \frac{2m_e}{\hbar^2} \langle \bs k^\prime | V | \Psi \rangle.
\end{align}
The proportionality constant $N$ depends on the normalization used for $|\bs k^\prime\rangle$ and $|\Phi \rangle$, and the plane wave bra $\langle \bs k^\prime |$ is defined from the position representation of the $\delta$-normalized plane wave ket $|\bs k\rangle$:
\begin{equation} \label{eq:normalized_plane_wave}
\langle \bs r | \bs k \rangle = \frac{\e^{\ii \bs k \cdot \bs r}}{(2\pi)^{3/2}}.
\end{equation}
To lowest order in $V$ (the first order Born approximation), the scattering amplitude is given by
\begin{align} \label{eq:born_scattering_amplitude}
 f^B[\Phi] &= - \frac{1}{4\pi} \frac{2m_e}{\hbar^2} \int \d^3 \bs r~\e^{-\ii \bs k^\prime \cdot \bs r} V(\bs r) \Phi(\bs r) \notag \\
 &= - \frac{N}{4\pi} \frac{2m_e}{\hbar^2} \langle \bs k^\prime | V | \Phi \rangle.
\end{align}
This describes a complex modulation of the spherical outgoing wave in eq.~\eqref{eq:scattering_wave_function} emanating from the scattering center.
This modulation is dependent on the function $|\Phi\rangle$, which represents the unperturbed incoming wave.

The formalism allows a direct definition of the differential scattering cross section as
\begin{equation} \label{eq:cross_section}
 \frac{\d\sigma}{\d\Omega} = |f(\bs k^\prime, \bs k)|^2.
\end{equation}
The scattering angle $\theta$ is the angle between the forward $z$ direction and the point of observation, such that:
\begin{equation} \label{eq:scattering_angle_momentum}
 \begin{aligned}
  &k_z^\prime = k \cos{\theta}, &k_\perp^\prime = k \sin{\theta}.
 \end{aligned}
\end{equation}
The quantity in eq.~\eqref{eq:cross_section} is the probability of detecting a scattered particle in the solid angle interval $[\Omega, \Omega+\d \Omega]$ for each incoming particle.
Thus, the modulus squared of the scattering amplitude is a directly measurable quantity.

An important observation to make is that the state $|\Phi\rangle$ has remained unspecified in the above formulas.
To obtain the Rutherford scattering amplitude, one must use a plane wave, but the popularity of this choice does not exclude more complicated and interesting forms of $|\Phi\rangle$.
As will be shown in section~\ref{sec:vortex_waves}, a Bessel beam can serve as a coherent generalization of a simple plane wave to cylindrically symmetric incoming waves.
To be able to compare the results obtained from a more general calculation in section~\ref{sec:vortex_born}, the simple plane wave case is given below.

The scattering amplitude for the screened Coulomb interaction potential (also called the Yukawa potential),
\begin{equation} \label{eq:screened_coulomb_potential}
 V(\bs r) = V(r) = V_0 \frac{\e^{-\mu r}}{r},
\end{equation}
can be calculated directly.
Using a $\delta$-normalized plane wave for $|\Phi\rangle = |\bs k \rangle$ (see eq.~\eqref{eq:normalized_plane_wave}), and filling in eq.~\eqref{eq:born_scattering_amplitude} with $N=(2\pi)^3$, we obtain the well-known result:
\begin{equation} \label{eq:rutherford_scattering_amplitude}
 f^B(k,\theta) = -\frac{2 m_e V_0}{\hbar^2} \frac{1}{4k^2\sin^2{\theta/2}+\mu^2},
\end{equation}
where $k$ is the conserved momentum and $\theta$ the scattering angle with respect to the incoming particle direction (here this is the $z$ axis).
The first order scattering amplitude is real, and no phase shifts can be directly inferred from this result.
Higher order terms can be calculated directly (which is often difficult) or by applying certain properties of the various terms in the Born series relating the real and imaginary parts of subsequently higher order terms~\cite{Zeitler}.
We do not consider higher order effects here, leaving this as a consideration for future work.
Care must also be taken if the screening parameter is set to zero, because then higher order terms in the Born series do not give the correct form of the asymptotic wave function, due to the long-range nature of the bare Coulomb potential~\cite{Holstein}.
Our current concern is an electron scattering from an atom, which is always screened to some extent, allowing use of the simple and uncorrected asymptotic form of eq.~\eqref{eq:scattering_wave_function}.

In the next section, the meaning and significance of Bessel beams as cylindrical basis states is illustrated by demonstrating how an aperture creates a localized superposition of Bessel states, and that Laguerre-Gaussian modes can be expanded in this basis.

\section{Vortex waves} \label{sec:vortex_waves}

\subsection{Bessel beams as plane waves}

The formal solutions to the field-free Schr\"odinger equation in cylindrical coordinates are Bessel states:
\begin{equation} \label{eq:bessel_beam}
 \langle \bs r | \bs k, \ell \rangle = \frac{\e^{\ii k_z z}}{\sqrt{2\pi}} J_\ell(\kappa r) \frac{\e^{\ii \ell\varphi}}{\sqrt{2\pi}},
\end{equation}
where $\kappa$ is the size of the transverse momentum.
The energy $E=\hbar^2(k_z^2+\kappa^2)/(2m_e)$ is independent of the OAM, $\hbar \ell$.
Physical insight into the structure of these solutions can be obtained by expanding eq.~\eqref{eq:bessel_beam} in a plane wave basis~\cite{Ivanov,Jentschura}:
\begin{subequations} \label{eq:bessel_plane_wave_expansion}
\begin{align}
 \langle \bs r|\bs k, \ell\rangle &= \e^{\ii k_z z} \int\frac{\d^2\bs k_\perp}{(2\pi)^2}~a_{\kappa,\ell}(\bs k_\perp) e^{\ii\bs k_\perp\cdot\bs r_\perp}, \\
 a_{\kappa,\ell}(\bs k_\perp) &= (-\ii)^\ell \e^{\ii\ell\phi} \frac{\delta(k_\perp-\kappa)}{\kappa}. \label{eq:vortex_fourier_coefficient}
\end{align}
\end{subequations}
A Bessel beam is thus nothing more than a coherent superposition of tilted plane waves which make an angle, $\theta_0 = \tan^{-1}(k_z/\kappa)$, with the $z$ axis, each with a $\varphi$-dependent phase.
This phase leads to the characteristic vortex beam OAM defined by the operator $\hat{L}_z = -\ii\hbar\pd_\varphi$ with eigenvalue $\hbar \ell$.
The important quantum numbers of the solutions are the energy $E$, the longitudinal momentum $k_z$, the transverse momentum $\kappa$, and the OAM index $\ell$.

Bessel beams form a complete set of solutions which can be used to compose more complicated beam profiles, such as those that occur in real experiments.
They play the role of ``vortex plane waves" for cylindrically symmetric wave functions, which is why they are used here for the calculation of the vortex wave scattering amplitude.

\subsection{Transversely localized beams} \label{sec:transverse_localization}

\begin{figure}
 \centering
 \includegraphics[width=.65\linewidth]{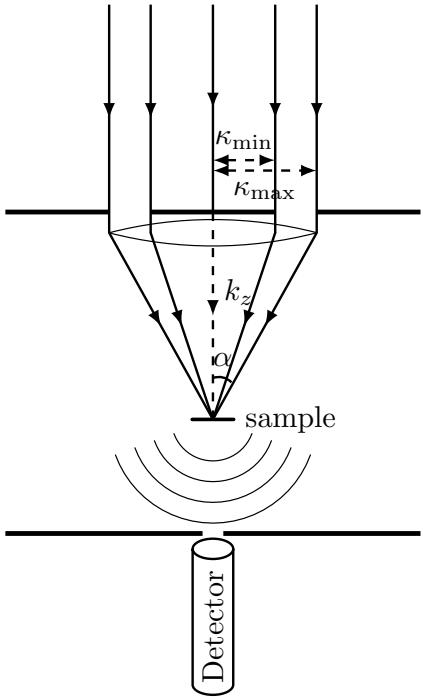}
 \caption{Schematic of the convergent beam scattering experiment. The relation of the transverse momenta and the aperture dimensions is shown, and an on-axis pinhole detector is shown. See sec.~\ref{sec:scattering_experiment} for the details of such an experiment. \label{fig:convergence_angle} \label{fig:scattering_experiment}}
\end{figure}

Bessel beams, as given in eq.~\eqref{eq:bessel_beam}, are a purely theoretical entity: they are not normalized in the transverse plane and therefore contain an infinite energy if one integrates over the spatial coordinates.
This is not necessarily a problem though, as for example plane waves have exactly the same issue, and this can be worked around by taking a superposition of different momenta, producing finite wave packets.
One must take care in calculating observables using only these basis states, as they can present misleading results when comparing to experiment.
A general method to localize a function in the transverse plane is to introduce a square integrable weighting function, $g(\kappa)$, that weights the contribution of various (transverse) momentum components, such that $\int_0^\infty \d^2\bs r_\perp~\Psi^*(\bs r_\perp)\Psi(\bs r_\perp) = 1$.
The resulting localized wave can then be written as follows:
\begin{equation} \label{eq:bessel_decomposition}
\Psi(r) = \int_0^\infty \d\kappa~\kappa~g(\kappa) \langle \bs r | \kappa, \ell \rangle.
\end{equation}
The simplest example of such a weighting function is a uniformly illuminated circular or annular aperture, for which $g(\kappa)$ is a step function over the aperture radius.
The minimum and maximum transverse momentum contributions are then directly related to the inner and outer radius of the aperture (in Fourier space), respectively, as shown in fig.~\ref{fig:convergence_angle}.
The total wave function in the far field of such an aperture is then:
\begin{equation} \label{eq:aperture_bessel_superposition}
 \Psi(r) \propto \int_{\kappa_\mathrm{min}}^{\kappa_\mathrm{max}} \d\kappa~\kappa~J_0(\kappa r).
\end{equation}
This has a simple interpretation: each infinitesimal ring of the aperture contributes exactly one Bessel beam with a precise transverse momentum.
For a small convergence angle $\alpha$, the maximally achievable transverse momentum component in the probe wave function (for $\alpha \ll 1$) is approximately equal to $k_z \alpha$.
\\

A Laguerre-Gaussian beam is a simple vortex beam model that is often used in optics as it is a natural laser mode.
It is a solution of the paraxial wave equation.
The normalized transverse wave function in the waist plane is given by:
\begin{align} \label{eq:laguerre_gaussian_beam}
 \langle \bs r | \ell, n, w \rangle =&~\Psi_n^\ell(r,\varphi) \notag \\
 =&~\frac{N}{w} \e^{\ii\ell\varphi} \left(\frac{r\sqrt{2}}{w}\right)^{|\ell|} L_n^{|\ell|} \left(\frac{2r^2}{w^2}\right) \exp{\left(-\frac{2r^2}{w^2}\right)}.
\end{align}
Here, $\ell$ is the OAM index, $n$ is a second index relating to the transverse localization of the mode (and the number of zero crossings of the Laguerre polynomial), $w$ is the beam waist, and the normalization is given by $N = \sqrt{\frac{2}{\pi}\frac{n!}{(n+|\ell|)!}}$.

Consulting mathematical literature~\cite{Szego}, one finds a useful relation between Bessel functions and the normalizable Laguerre-Gaussian modes, by way of a specific weighting function:
\begin{align}
 \Psi_n^\ell(r,\varphi) =&~\frac{N}{w} \e^{\ii \ell\varphi} \frac{1}{n!} \notag \\
 &\times \int_0^\infty \d \kappa~\frac{\e^{-\kappa^2 w^2 / 8}}{\sqrt{2}} \left(\frac{\kappa w}{\sqrt{8}}\right)^{2n+\ell+1}  J_\ell(\kappa r).
\end{align}
It is clear from the above that a Laguerre-Gaussian beam profile can be written as a Gaussian distribution of Bessel modes.
The weighting function $g(\kappa)$, (see eq.~\eqref{eq:bessel_decomposition}) of a normalized Laguerre-Gaussian at the waist plane into the Bessel basis can be written explicitly as:
\begin{equation}
 g(\kappa) = \frac{1}{w} \sqrt{\frac{2\times n!}{(n+\ell)!}} \frac{1}{n!} \frac{1}{\kappa} \left(\frac{\kappa w}{2}\right)^{2n+\ell+1} \exp{\left(-\frac{\kappa^2 w^2}{4}\right)}.
\end{equation}

\begin{figure}
\centering
\includegraphics[width=\linewidth]{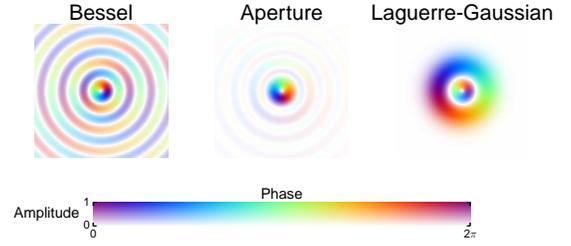}
\caption{(Color online) Transverse wave functions for the Bessel (eq.~\eqref{eq:bessel_beam}), aperture far field, and Laguerre-Gaussian (eq.~\eqref{eq:laguerre_gaussian_beam} beams, using the HSV color map as shown in the legend at the bottom. They all contain an $\ell=1$ vortex. The Laguerre-Gaussian has two intensity lobes ($n=2$), clearly showing the strongest localization of the three in the transverse plane. \label{fig:vortex_transverse_profiles}}
\end{figure}

Figure~\ref{fig:vortex_transverse_profiles} compares the complex transverse profiles of a pure single-$\kappa$ Bessel mode, the far field of a circular aperture with uniform intensity and a centered phase vortex, and an $\ell=1$, $n=2$ Laguerre Gaussian mode. The latter is most strongly localized, followed by the aperture far field.
The Bessel mode is much broader and not localized.
\\

The above establishes that Bessel beams can serve as a proper basis for beam states, through a weighted integral over $\kappa$.
A single Bessel component contains a coherent circular superposition of transverse momenta, with an azimuthal phase relation related to the order of the Bessel function.
We can now use this basis to calculate a generalized Coulomb scattering amplitude for Bessel beams.

\section{Vortex Born scattering amplitude} \label{sec:vortex_born}

\subsection{Fourier expansion}

The first order Coulomb scattering amplitude for a plane wave, eq.~\eqref{eq:rutherford_scattering_amplitude} is a well-known result.
In this section, the incoming wave $|\Phi\rangle$ will not be taken to be a plane wave, but instead we use a Bessel beam as in eq.~\eqref{eq:bessel_beam}.
In essence, the modulating factor of a spherical plane wave emanating from the scattering center will be calculated for an incoming Bessel beam.
The first step is to apply eq.~\eqref{eq:bessel_plane_wave_expansion}.
Through this substitution the integral over $\bs r$ can be performed resulting in the plane wave scattering amplitude result, leaving the two integrations over $\bs k_\perp$ introduced by eq.~\eqref{eq:bessel_plane_wave_expansion}.
The factor $N$ is chosen to be $(2\pi)^{5/2}$ here, canceling the normalization factors of the initial and final states as in the plane wave case:
\begin{align}
 f_V^B =& -\frac{N m_e}{2\pi\hbar^2} \langle \bs k^\prime|V|\bs k, \ell\rangle \notag \\
 =& -\frac{m_e}{(2\pi)^2\hbar^2} \int \d^3\bs r~\e^{-\ii\bs k^\prime\cdot \bs r} V(\bs r) \notag \\
 &\times\int\d^2\bs k_\perp~(-\ii)^\ell \e^{\ii\ell\phi} \frac{\delta(k_\perp-\kappa)}{\kappa} \e^{\ii (\bs k_\perp \cdot \bs r_\perp} \e^{\ii k_z z)} \notag
\end{align}
\begin{align}
 f_V^B =& -\frac{m_e (-\ii)^\ell}{(2\pi)^2\hbar^2\kappa} \int \d^2\bs k_\perp~e^{\ii \ell\phi} \delta(k_\perp - \kappa) \notag \\
  &\times \int \d^3\bs r~e^{\ii\bs q \cdot \bs r} V(\bs r)\notag \\
  =& -\frac{m_e (-\ii)^\ell V_0}{\pi\hbar^2 \kappa} \int_0^\infty \d k_\perp \int_0^{2\pi} \d\phi~k_\perp \frac{\delta(k_\perp - \kappa) \e^{\ii\ell\phi}}{q^2+\mu^2} \notag \\
 =&-\frac{m_e (-\ii)^\ell V_0}{\pi\hbar^2} \notag \\
 &\times \int_0^{2\pi}\d\phi~\frac{e^{\ii\ell\phi}}{(\kappa \cos{\phi} - k_x^\prime)^2 + (\kappa \sin{\phi} - k_y^\prime)^2 + \chi^2},
\end{align}
where $\chi^2 = q_z^2 + \mu^2$ was used.
In the last step, the integration over $k_\perp$ was performed, bringing us to the main part of the calculation.
The vector $\bs k^\prime$ is the wave vector of the scattered plane wave.

\subsection{Contour integration}

\begin{figure}
 \centering
 \includegraphics[width=.75\linewidth]{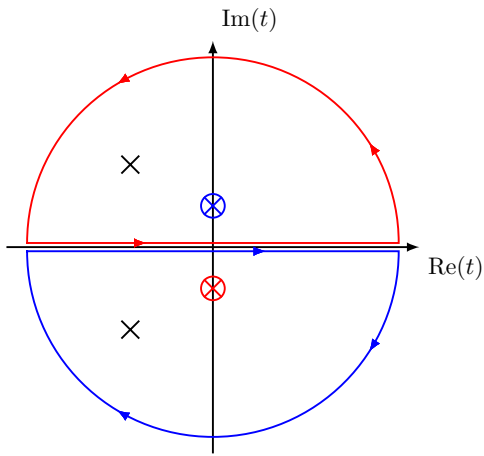}
 \caption{(Color online) Poles and contours of the integrand in eq.~\eqref{eq:vortex_born_trigonometric}. The red (top) contour is used for $\ell>0$. The blue (bottom) contour is used for $\ell<0$. In both cases, there is an $\ell$-th order pole in the other half of the complex plane due to the first factor of eq.~\eqref{eq:vortex_born_trigonometric}. Only one pole of those due to the second factor in eq.~\eqref{eq:vortex_born_trigonometric} lies in a contour for every value for $\ell$. \label{fig:poles}}
\end{figure}

To be able to integrate over the Fourier angular coordinate, we perform the following substitution:
\begin{equation}
 t = \tan{\frac{\phi-\pi}{2}}.
\end{equation}
This results in an integration over all real values of $t$, which we plan to perform in turn by means of a semicircular contour in the complex plane over $t$:
\begin{widetext}
\begin{equation} \label{eq:vortex_born_trigonometric}
 f_V^B = -\frac{m_e \ii^\ell V_0}{\pi\hbar^2} \oint_{\Gamma_\ell} \left(\frac{\ii-t}{\ii+t}\right)^\ell \frac{\d t}{\left[\left((\kappa-k_x^\prime)^2 + {k_y^\prime}^2 + \chi^2\right)t^2 + 4k_y^\prime\kappa~t + (\kappa+k_x^\prime)^2 + {k_y^\prime}^2 + \chi^2\right]},
\end{equation}
The contour $\Gamma_\ell$ depends on the sign of $\ell$, as discussed below and shown in fig.~\ref{fig:poles}.
We have reduced the calculation of the scattering amplitude to determining the right contour and calculating the residues of the poles inside that contour.

The first factor's poles at $+\ii$ or $-\ii$, lie either above or below the real axis.
The multiplicity of these poles and thus the difficulty in calculating their residue increase with increasing $\ell$.
Consequently it is beneficial to keep these outside the contour.
Intuitively, we can predict that the complexity of the solution should be independent of the magnitude of $\ell$, so we shall choose the semi-circle contour $\Gamma_\ell$ such that these ``vortex poles" lie \emph{outside} the contour, and thus, do not contribute to the integral.
The denominator of the second factor gives the two remaining poles:
\begin{equation} \label{eq:vortex_born_trigonometric_poles}
 t_\pm = \frac{-2k_y^\prime \kappa \pm \sqrt{2({k_\perp^\prime}^2-\chi^2)\kappa^2 - \kappa^4 - ({k_\perp^\prime}^2 + \chi^2 )^2}}{(k_x^\prime-\kappa)^2 + {k_y^\prime}^2 + \chi^2}.
\end{equation}
It can be shown (see the Appendix) that the square root is a pure imaginary number, \textit{i.e.}
\begin{equation} \label{eq:vortex_born_trigonometric_sqrt}
 2({k_\perp^\prime}^2 - \chi^2)\kappa^2 < \kappa^4 + ({k_\perp^\prime}^2 + \chi^2)^2.
\end{equation}
This means that only one pole will lie inside the contour for a given sign of $\ell$ (with the other lying on the opposite side of the real axis).
Figure~\ref{fig:poles} shows a sketch of the contour and location of the poles.
By calculating the residues for both signs of $\ell$, one notices the result only depends on the sign of $\ell$ through a phase factor $\e^{\ii \ell \phi^\prime}$, resulting in a residue of the pole inside the contour $\Gamma_\ell$ equal to:
\begin{equation}
 \res_{t=t_\pm} = 
 \frac{\e^{\ii \ell \phi^\prime}}{2 \sqrt{2 ({k_\perp^\prime}^2 - \chi^2) \kappa^2 - \kappa^4 - ({k_\perp^\prime}^2 + \chi^2)^2}} \left(\frac{{k_\perp^\prime}^2 + \chi^2 + \kappa ^2 - \sqrt{\left({k_\perp^\prime}^2 + \chi^2\right)^2+\kappa^4-2 \left({k_\perp^\prime}^2 - \chi^2\right) \kappa^2}}{-2\kappa k_\perp^\prime}\right)^{|\ell|}.
\end{equation}
The final result for the scattering amplitude is then obtained by applying the residue theorem. This is the end of the scattering amplitude calculation, and the final result can summarized as:
\begin{equation} \label{eq:vortex_scattering_amplitude}
\begin{aligned}
 f_V^B =& -\frac{2 m_e  V_0}{\hbar^2}~ \ii^\ell \e^{\ii\ell\phi^\prime} \left(\frac{{k_\perp^\prime}^2 + \chi^2 + \kappa ^2 -  \sqrt{\left({k_\perp^\prime}^2 + \chi^2\right)^2+\kappa^4-2 \left({k_\perp^\prime}^2 - \chi^2\right) \kappa^2}}{-2\kappa k_\perp^\prime}\right)^{|\ell|} \\
 &\times \frac{1}{\sqrt{\kappa^4 + ({k_\perp^\prime}^2 + \chi^2)^2 - 2({k_\perp^\prime}^2 - \chi^2)\kappa^2}}.
\end{aligned}
\end{equation}
\end{widetext}

\subsection{Discussion} \label{sec:discussion}

Although the expression in eq.~\eqref{eq:vortex_scattering_amplitude} is not simple, there are three key constituents, which can each be attributed to the physical concepts that are important in this system.
The first is the vortex phase factor $\e^{\ii \ell \phi^\prime}$, which signifies the expected conservation of OAM.
The second factor, $(\hdots)^{|\ell|}$, is the contribution due to the order of the Bessel function (and thus indirectly the OAM of the beam).
Note that this contribution is independent of the sign of $\ell$, as expected.

\begin{figure}
\centering
 \includegraphics[width=\linewidth]{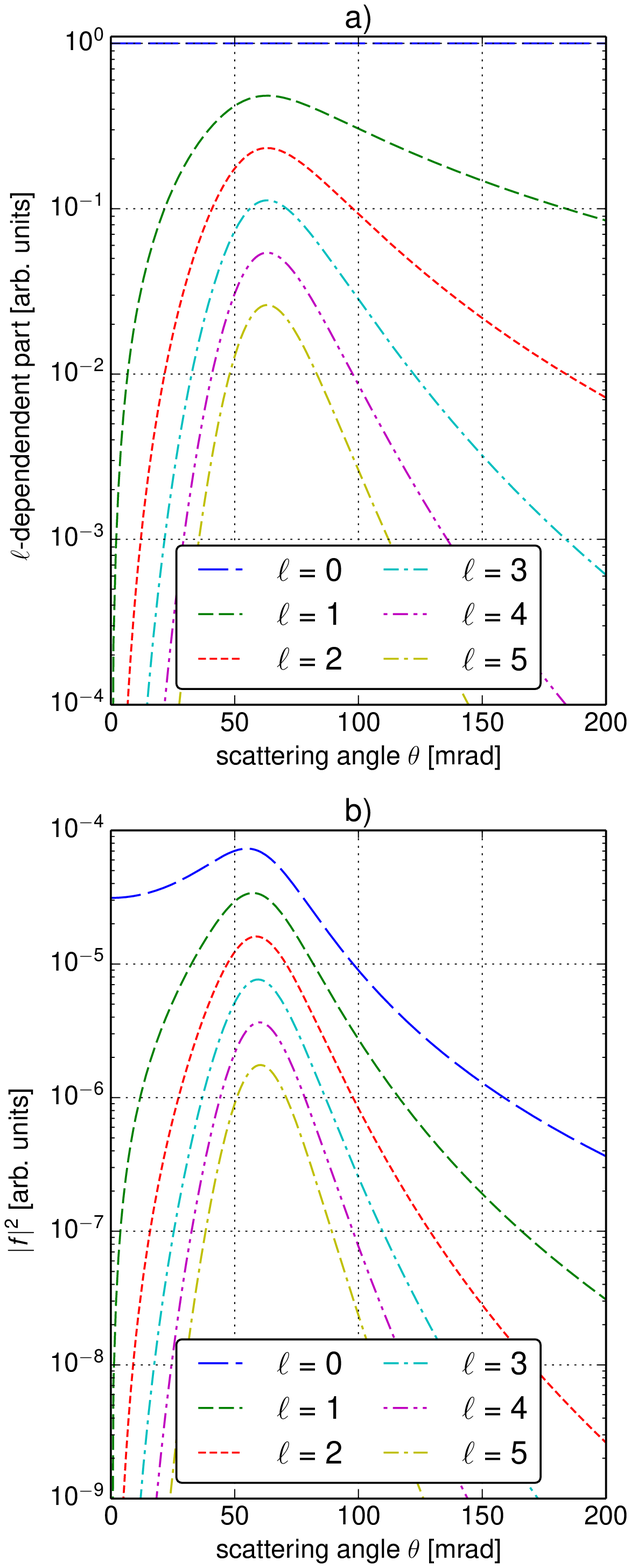}
 \caption{(Color online) The elastic scattering cross section for a screened Coulomb potential and Bessel beam for various values of $\ell$, $\kappa=25~{a_0}^{-1}$, $Z=26$, and $k_z=169~{a_0}^{-1}$ (corresponding to a 300~kV acceleration voltage). See Sec.~\ref{sec:discussion} for details on the screening potential model for an atom with atomic number $Z$ used. The scattering amplitude becomes smaller with larger $\ell$. a) only the $(\hdots)^{|\ell|}$ factor. b) the full equation~\eqref{eq:vortex_scattering_amplitude}. \label{fig:ell_dependence}}
\end{figure}

\begin{figure}
\centering
 \includegraphics[width=\linewidth]{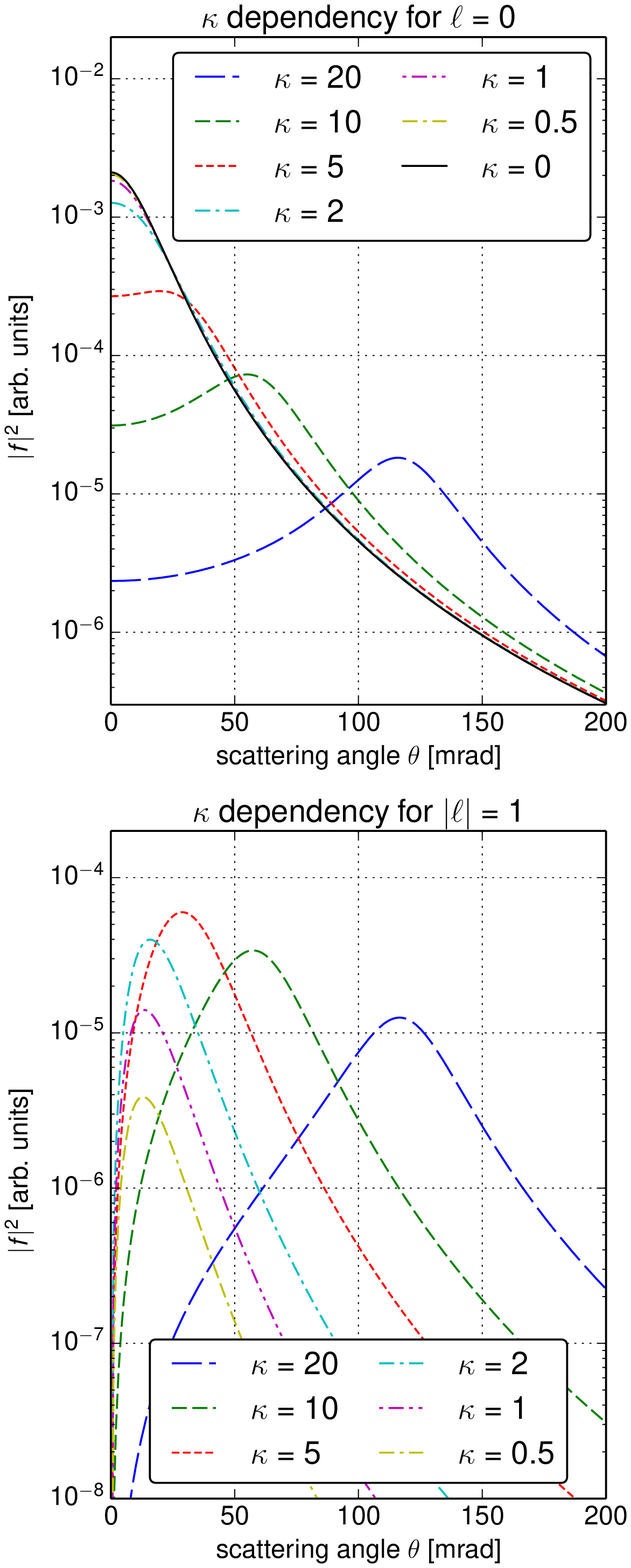}
 \caption{(Color online) The elastic scattering cross section for a screened Coulomb potential and Bessel beam for various values of $\kappa$ (in ${a_0}^{-1}$), $\ell=0,\pm1$, $Z=26$, and $k_z=169~{a_0}^{-1}$ (corresponding to a 300~kV acceleration voltage). See Sec.~\ref{sec:discussion} for details on the screening potential model for an atom with atomic number $Z$ used. The scattering amplitude becomes smaller with larger $\ell$, due to the larger spatial spread of higher-$\ell$  Bessel functions. \label{fig:kappa_dependence}}
\end{figure}

The $(\hdots)^{|\ell|}$ function is shown in fig.~\ref{fig:ell_dependence}a.
This independence of the sign of $\ell$ was assumed nowhere in the calculation (both signs were considered separately in light of the integrand's poles as shown in fig.~\ref{fig:poles}).
The third and last factor is the zeroth order Bessel contribution, which as will be shown below, reduces to the plane wave result in the appropriate limit.
Figures~\ref{fig:ell_dependence}b and~\ref{fig:kappa_dependence} show the differential scattering cross section $|f|^2$ for various values of $\ell$ and $\kappa$.

The figures were made using the improved Hartree-Fock and Thomas-Fermi-Dirac atomic screening model~\cite{Jablonski}.
In this model, the atomic screening factor becomes a (numerical) nonlinear function of the atomic number $Z$:
\begin{equation}
\mu(Z) = \mu_\infty(Z) \frac{Z^{1/3}}{0.8853},
\end{equation}
where $\mu_\infty(Z)$ is calculated numerically from atomic wave functions and is tabulated in ref.~\cite{Jablonski}.

The most important feature of the scattering amplitude for vortex $|\ell|>0$ waves is the zero intensity at the center, which together with the azimuthal phase factor (see eq.~\eqref{eq:vortex_scattering_amplitude}) results in a vortex wave emitted from the scattering center, as expected.
Due to the symmetry of the system, no OAM is lost or gained, and the vorticity is not altered.
It is also found that for the same value of $\kappa$, higher $\ell$ beams scatter less.

This is due to the increase of radius with $\ell$ (for the same $\kappa$), leading to a reduction of the overlap with the potential (eq.~\eqref{eq:screened_coulomb_potential}) and therefore, less interaction.
This has also been observed in electron channeling calculations~\cite{Xin}.

One can clearly see the plane wave limit take shape in fig.~\ref{fig:kappa_dependence}a as $\kappa\rightarrow 0$.
For high $\kappa$ input beams, a novel off-axis $\ell=0$ peak appears.
It might be interesting to investigate if this peak can be detected in electron-atom scattering experiments.
The value $\kappa$ for which this peak starts to appear is precisely $\mu$, the value of the screening parameter in the screened Coulomb potential (see eq.~\eqref{eq:screened_coulomb_potential}).
The implications of this for the experimental observation of this peak are discussed after the plane wave limit below.

\subsection{Plane wave limit}

We now prove that eq.~\eqref{eq:vortex_scattering_amplitude} calculated here is a proper generalization of the well-known Rutherford scattering formula given by eq.~\eqref{eq:born_scattering_amplitude}, by taking the appropriate plane wave limit.
To take a Bessel beam to its plane wave limit, two limits are taken: $\ell=0$, and $\kappa\rightarrow 0$.
The first is trivial, and the second can be intuitively made plausible by considering the zeroth order Bessel function $J_0(\kappa r)$.
As its parameter $\kappa$ becomes very small, the period of the oscillatory behavior of the function becomes infinitely long, up to the point the function approaches a constant value of 1.
Thus, for $\kappa\rightarrow 0$, a Bessel beam as in eq.~\eqref{eq:bessel_beam} of order 0 becomes a plane wave traveling in the $z$-direction.
Note that for $\ell\neq0$ and $\kappa\rightarrow0$, the Bessel beam solution becomes zero everywhere, signifying that there can be no OAM without transverse momentum.

To be able to easily compare eq.~\eqref{eq:vortex_scattering_amplitude} to the usual elastic scattering amplitude, it is valuable to substitute some unusual variables for the conserved size of the momentum $k$ and the scattering angle $\theta$ using eq.~\eqref{eq:scattering_angle_momentum} and:
\begin{equation}
 q_z^2 = (k_z-k_z^\prime)^2 = (k_z - k \cos{\theta})^2.
\end{equation}
Expanding eq.~\eqref{eq:vortex_scattering_amplitude} up to first order in $\kappa$, and setting $\ell$ to 0, one obtains:
\begin{align}
 f_V^B &\approx -\frac{2 m_e V_0}{\hbar^2} \frac{1}{2k_z^2(1-\cos{\theta})+\mu^2} + \mathcal{O}(\kappa^2).
\intertext{This can be rewritten as}
 f_V^B &\approx -\frac{2 m_e V_0}{\hbar^2} \frac{1}{4 k_z^2 \sin^2{(\theta/2)} + \mu^2},
\end{align}
which agrees exactly with the normal Rutherford scattering result, and confirms the choice of $N=(2\pi)^{5/2}$ to be correct.
This concludes the reduction of our result to that for plane waves, proving the result presented in eq.~\eqref{eq:vortex_scattering_amplitude} is in fact far more general.
It is important that the implications of the transverse momentum as a new degree of freedom are investigated further.

\subsection{Zeroth order off-center peak} \label{sec:peak_observability}

A distinct feature of eq.~\eqref{eq:vortex_scattering_amplitude} for $\ell=0$ is shown clearly in fig.~\ref{fig:kappa_dependence}.
For high $\kappa$, the maximum shifts away from the center, producing a ring-like scattering amplitude instead of a centered peak.
This is a result of incorporating larger transverse momenta in the scattering process.
It must be noted that this is a purely elastic effect, and the authors believe it has not been discussed or observed before, due to the exotic circumstances (extremely high convergence angles) required.
This effect is also present for nonzero values of $\ell$, although it is much less pronounced (see for example the slightly linear region of the $\kappa=20~{a_0}^{-1}$ curve in fig.~\ref{fig:kappa_dependence}).

The shift of intensity to the off-center peak makes the $\theta=0$ scattering amplitude decrease with increasing $\kappa$.
Filling in $k_\perp^\prime=0$ in eq.~\eqref{eq:vortex_scattering_amplitude} gives this central scattering amplitude:
\begin{equation}
f_V^B(k~\bs e_z; \bs k, 0) = -\frac{2m_e V_0}{\hbar^2} \frac{1}{(k-k_z)^2+\kappa^2+\mu^2},
\end{equation}
a monotonically decreasing function of the original transverse momentum $\kappa$.
This is expected from fig.~\ref{fig:kappa_dependence}: a higher transverse momentum scatters more electrons away from the center.
This effect is unique to a description which incorporates transverse momentum.

\subsection{Off-center vortex Born scattering amplitude}

While plane waves have no preferred center of symmetry, cylindrical vortex waves posses a phase singularity at a specific point.
The relative displacement of the scattering center and this symmetry axis influences the OAM spectrum of the incoming beam as seen by the scatterer.
This effect has been discussed extensively and measurement of the OAM of a displaced beam will reveal an \emph{OAM spread} around the classical \emph{mean OAM} value~\cite{Vasnetsov,Zambrini}, defined as $\int \d^3 \bs r~ \psi^* \hat{L}_z \psi$.
In quantum mechanics, this is the expectation value of the OAM operator, which is not necessarily the value one would measure, but rather the statistical average of a series of measurements.
This is why a beam's OAM is \emph{quasi-intrinsic}~\cite{Zambrini}.
Displacing a beam results in a superposition of OAM eigenstates impinging on the scattering center, which will influence the selection rules for inelastic scattering~\cite{Picon,Yuan}.

As applied recently in literature~\cite{Yuan,Afanasev}, the Bessel addition theorem~\cite{Abramowitz,Watson} allows a simple way to calculate off-center scattering contributions.
A Bessel beam with its center displaced by a vector $\bs{r_{0_\perp}}$, can be expressed as follows:
\begin{equation} \label{eq:bessel_displacement}
 J_\nu(\kappa r^\prime) \e^{\ii \ell \varphi^\prime} = \sum_{m=-\infty}^{+\infty} J_{\ell+m}(\kappa r) \e^{\ii(\ell+m)\varphi} J_m(\kappa r_0) \e^{-\ii m\varphi_0}.
\end{equation}
This formula is a mathematical identity that reveals quite interesting properties.
As is already clear from misaligned vortex OAM analysis~\cite{Vasnetsov} and reflected in eq.~\eqref{eq:bessel_displacement}, the mean OAM is always equal to $\hbar\ell$.
The formula also shows the size of each contribution is modulated by the value $J_m(\kappa r_0)$.
For large $r_0$, this function goes to zero.
For large $m$ and small $r_0$, it also tends to zero.
The OAM spread is limited by these factors, but in a physical beam, where there is a necessary integration over a region of $\kappa$ values (as discussed in section~\ref{sec:vortex_waves}), the spread is expected to increase.
In the $\kappa\rightarrow0$ limit, eq.~\eqref{eq:bessel_displacement} ensures only the zeroth order contributes as it should (a plane wave has no preferred center).
This theorem will now be applied to give a simple expression for off-center elastic scattering.

The off-center incoming wave in eq.~\eqref{eq:bessel_displacement} can be expressed in ket notation as follows:
\begin{equation}
 |\bs k, \ell, \bs r_{\perp0} \rangle = \sum_{m=-\infty}^{+\infty} \e^{-\ii m \varphi_0} J_m(\kappa r_0) |\bs k, \ell+m \rangle,
\end{equation}
The off-center elastic contribution can be directly written down as:
\begin{align}
 f_{V, \bs r_0}^B &= -\frac{m_e}{2\pi\hbar^2} \langle \bs k^\prime | V | \sum_{m=-\infty}^{+\infty} \e^{-\ii m \varphi_0} J_m(\kappa r_0) |\bs k, \ell+m\rangle \notag \\
 &= \sum_{m=-\infty}^{+\infty} \e^{-\ii m \varphi_0} J_m(\kappa r_0) f^B_V (\bs k^\prime; \bs k, \ell + m). \label{eq:vortex_off_center_scattering_amplitude}
\end{align}
If one is interested only in the $\theta=0$ scattered intensity, one can set $k_\perp^\prime=0$ (all momentum in the forward direction), which eliminates the sum over $m$ through $f_V^B$, which is only nonzero in this case if $\ell+m=0$:
\begin{equation}
 f_{V, \bs r_0}^B(k~\bs e_z; \bs k, \ell) = (-1)^\ell \e^{\ii \ell \varphi_0} J_\ell(\kappa r_0) f^B_V (k~\bs e_z; \bs k, 0).
\end{equation}
This means that for an off-center vortex scattering on an atom, the scattering amplitude in the forward direction is equal to the scattering amplitude of a centered non-vortex wave weighted by $J_\ell(\kappa r_0)$.
Note that for physically realistic beams, this needs to be integrated over the relevant $\kappa$ range, effectively blurring the amplitude over these momentum components.

Even for $\ell=0$, non-zero OAM modes contribute coherently to the scattering amplitude, although the resulting mean OAM is zero as required.
The mean OAM is also conserved in the scattering process.
The $\ell=0$ contribution, which scatters the most strongly, is largest for displacements with a maximum of the incoming Bessel function $J_\ell$ on top of the scattering center.

When a properly localized electron vortex beam excites an internal atomic state inelastically, the vortex phase factor $\e^{\ii \ell\varphi}$ will alter the usual selection rules and allow normally forbidden transitions to higher angular momentum states.~\cite{Lloyd}
The sum in eq.~\eqref{eq:vortex_off_center_scattering_amplitude} means that an off-center beam will excite a variety of transitions, with weights related to the displacement distance and the angular momenta involved in the transition.~\cite{Yuan}
These unwanted contributions can theoretically be made arbitrarily small by careful alignment.

The mean OAM of a displaced beam stays constant under parallel displacement, and each OAM component scatters independently and unitarily, so the mean OAM of the scattered wave remains unaltered by elastic scattering, unlike the inelastic case.
One might expect some rotational motion to be transferred to the atom, but there was no recoil taken into account here.
Relaxing the approximations made by including recoil of the scattering center (\textit{e.g.} an atom) will allow for an accurate calculation of the elastic momentum transfer.
This transfer of momentum has already been observed in experiments~\cite{Verbeeck_tweezers}, and is thus expected to appear in more detailed approaches.

\section{Proposal for new dark-field imaging method} \label{sec:scattering_experiment}

The theory developed in this paper can be directly applied to describe a new dark-field imaging method.
For $\ell\neq0$ beams the central dark spot's size is described accurately by eq.~\eqref{eq:vortex_scattering_amplitude}, perhaps employing eq.~\eqref{eq:vortex_off_center_scattering_amplitude} to account for order mixing due to misalignment.
It is in this dark spot that evidence of OAM transfer of the electron vortex to the atomic system can be detected.
Because this is a dark region in the scattered intensity, it is advantageous to measure here due to counting statistics.
A smaller intensity will increase the signal-to-noise ratio, and ensure that even small OAM transfer effects can be detected.

The loss of OAM by the vortex electron has been described theoretically by various authors and can be caused by various mechanisms, \textit{e.g.} atomic transitions~\cite{Davis,Lloyd,Babiker,Bhattacharya,vanVeenendaal} or due to symmetry~\cite{Ferrando}.
The interaction with a vortex can induce novel transitions (due to the symmetry change induced by the vortex phase), which can be detected where the elastic scattering amplitude is small.
In a perfectly aligned electron-optical system, any electrons that end up in the central dark spot will have lost OAM.
In the $\ell\neq0$ case, no elastically scattered electrons can end up there.

The experimental setup is shown in fig.~\ref{fig:scattering_experiment}.
A convergent electron beam that can be tuned arbitrarily in $\kappa$ content by virtue of eq.~\eqref{eq:aperture_bessel_superposition}, impinges on an atomic system, scattering the electrons as described by eq.~\eqref{eq:vortex_scattering_amplitude} of which examples are shown in figs.~\ref{fig:ell_dependence} and~\ref{fig:kappa_dependence}.
Annular electron apertures, such as that used in \cite{Clark} can be used here.

The advantage of this imaging method is that only the electrons that are important to the interaction are detected, significantly increasing sensitivity to small signals.
This dark field measurement will enjoy an increased signal to noise ratio, allowing the detection of weak interactions.

\section{Conclusion}

In this paper, we considered elastic scattering of vortex electrons on a single atom, modeled by a screened Coulomb potential.
First, we showed how Bessel beams can be used as a complete basis for any physical beam profile and explained their quantum numbers.
To illustrate this property, an important class of beam solutions, the Laguerre-Gaussian modes (in the waist plane), were expanded into a distribution of Bessel beams over the transverse momentum.

The quantum theory of scattering was reviewed to derive the Rutherford scattering formula within the Born approximation.
This plays an important role in elastic atomic scattering, inelastic scattering, and backscattering experiments, and the results presented here are fundamental to develop and understand future vortex scattering experiments.
The usual Rutherford scattering amplitude was generalized by considering a general cylindrically symmetric incoming state, and an exact analytical expression is obtained for the first order elastic scattering amplitude.
The result was shown to successfully generalize the plane wave scattering amplitude to general cylindrically symmetric beams.
Novel features appear in this generalized description, warranting further investigation, both theoretically, and experimentally.
Off-axis scattering is discussed for Bessel beams, and the broadening effect of the OAM contributions is explained for elastic and inelastic scattering.

As an application of the theory developed in this paper, a new dark-field imaging technique using electron vortex beams is proposed.
The elastic scattering dark spot is used to enable the detection of inelastic signals with a much higher signal-to-noise ratio than currently available.
This gives a technique focused on measuring in the vortex' dark spot a significant advantage due to the improved counting statistics.

The first Born approximation is generally insufficient when the scattering involves more than an atom or molecule, although it serves as an important baseline that can be used for comparison of more advanced calculations.
These improvements include higher order Born terms, alternative formalisms such as eikonal scattering (and the associated high energy approximations), and dynamical theories such as that for electron channeling.
The results presented in this paper can be tested experimentally, provide necessary quantification to design future experiments, and give insight into electron vortex scattering that will prove valuable when developing a quantitative inelastic scattering theory and experiments.

\acknowledgments
RVB acknowledges support the FWO (Aspirant Fonds Wetenschappelijk Onderzoek
- Vlaanderen).
JV acknowledges financial support from the EU under the Seventh Framework Program (FP7) under a contract for an Integrated Infrastructure Initiative, Reference No. 312483-ESTEEM2, and the European Research Council under the FP7, ERC grant N246791 - COUNTATOMS and ERC Starting Grant 278510 VORTEX.

\appendix
\section{The pole square root}

We show here under what conditions
\begin{equation} \label{eq:proof_sign}
 ({k_\perp^\prime}^2 + \kappa^2 + \chi^2)^2 - c\kappa^2{k_\perp^\prime}^2 \geq 0,
\end{equation}
Where all variables are real numbers and $k_\perp^\prime, \kappa, \chi, c~>~0$.
We assume the above to be true, and consider the resulting condition:
\begin{equation} \label{eq:vortex_plane_sqrt}
 ({k_\perp^\prime}^2 + \kappa^2 + \chi^2)^2 - c\kappa^2{k_\perp^\prime}^2 \geq 0
\end{equation}

\begin{align}
 \Leftrightarrow \frac{{k_\perp^\prime}^2 + \kappa^2 +  \chi^2}{\sqrt{c}\kappa k_\perp^\prime} &\geq 1 \notag \\
 \Leftrightarrow -(\chi^2+(2-\sqrt{c})\kappa k_\perp^\prime) &\leq (\kappa - k_\perp^\prime)^2. \label{eq:proof_sign_general_condition}
\end{align}
Now, $\chi \in \mathbbm{R}$, which means $\chi^2\geq0$, thus $-\chi^2\leq0$.
This inequality is always true if $(2-\sqrt{c}) \geq 0$, which means eq.~\eqref{eq:proof_sign} is always true if $|c| \leq 4$.
In case $|c|>4$, the validity of eq.~\eqref{eq:proof_sign} reduces to eq.~\eqref{eq:proof_sign_general_condition}.

\bibliography{electron_vortex_coulomb_scattering}

\end{document}